\documentclass[prl,superscriptaddress,showpacs]{revtex4}

\usepackage{enumerate}
\usepackage{amsfonts,amssymb,amsmath}
\usepackage[]{graphics,graphicx,epsfig}
\usepackage{amsthm}
\usepackage{color}
\usepackage{tabularx}
\usepackage{array}
\begin{document}

\title{An experimental proposal for revealing contextuality in almost all qutrit states}

\author{Jayne \surname{Thompson}}
\affiliation{Centre for Quantum Technologies, National University of Singapore, 3 Science Drive 2, 117543 Singapore, Singapore}

\author{Robert \surname{Pisarczyk}}
\affiliation{Centre for Quantum Technologies, National University of Singapore, 3 Science Drive 2, 117543 Singapore, Singapore}
\affiliation{Department of Physics, University of Oxford, Oxford UK}

\author{Pawe{\l} \surname{Kurzy\'nski}}
\affiliation{Centre for Quantum Technologies, National University of Singapore, 3 Science Drive 2, 117543 Singapore, Singapore}
\affiliation{Faculty of Physics, Adam Mickiewicz University, Umultowska 85, 61-614 Pozna\'{n}, Poland}

\author{Dagomir \surname{Kaszlikowski}}
\email{phykd@nus.edu.sg}
\affiliation{Centre for Quantum Technologies, National University of Singapore, 3 Science Drive 2, 117543 Singapore, Singapore}
\affiliation{Department of Physics, National University of Singapore, 2 Science Drive 3, 117542 Singapore, Singapore}

\begin{abstract}
Contextuality is a foundational phenomenon underlying key differences between quantum theory and classical realistic descriptions of the world.
Here we propose an experimental test which is capable of revealing contextuality in all qutrit systems, except the completely mixed state, provided we choose the measurement basis appropriately. The 3-level system is furnished by the polarization and spatial degrees of freedom of a single photon, which encompass three orthogonal modes. Projective measurements along rays in the 3-dimensional Hilbert space are made by linear optical elements and detectors which are sensitive to single mode. We also discuss the impact of detector inefficiency and losses and review the theoretical foundations of this test.
\end{abstract}

\maketitle

\section{Introduction}

In classical physics, all observable quantities have an objective reality. The outcome of measuring a classical observable $A$ can not be influenced by a property of the observer or by the choice to simultaneously co-measured $A$ with a second observable $B$.  This property is referred to as non-contextual realism.

Distinctively, in quantum theories the measurement outcome for an observable $A$ depends on what you choose to co-measure at the same time. There exist certain sets of observables where it is impossible to simultaneously assign pre-existing outcomes. In this scenario there is no joint probability distribution from which the measurement statistics for every observable can be recovered as marginals.

Recently Klyachko, Can, Binicioglu and Shumovsky introduced a  5-ray (KCBS) inequality \cite{Klyachko}. The KCBS inequality bounds the maximum attainable level of correlation between outcomes of $5$ observables, $A_1,\dots, A_5$, under the dual assumptions that only $A_i$ and $A_{(i+ 1 \, {\rm mod } \, 5)}$ are co-measurable and that it is possible to preassign outcomes to all five observables in accordance with non-contextual realism \cite{Footnote1}. By construction the KCBS inequality is satisfied by all non-contextual hidden variable models. However Klyachko et. al. discovered violation of the inequality in a 3-level quantum system. Setting aside Bell type tests of nonlocality \cite{Footnote2}, it has been shown that the KCBS inequality is the simplest possible test for contextuality of quantum theories, in the sense that any inequality based on projective measurements along less than five different direction will be satisfied by all classical and quantum theories \cite{Ramanathan}. The KCBS inequality formed the basis of several recent experimental tests of quantum contextuality. Lapkiewicz et. al. have experimentally demonstrated violation of the KCBS inequality in an indivisible 3-level quantum system (furnished by a single photon) \cite{Zeilinger}. This experiment built upon a robust and rapidly developing body of experimental contextuality literature \cite{Kirchmair,Zu,Moussa,Bartosik,Amselem,Cabello}. In our present work we draw on the tests outlined in these papers.

From the perspective of an experimental test of non-contextual realism the KCBS inequality is a state dependent test; for a sufficiently pure state there exists a choice of five measurement directions which can be used to detect violation of the KCBS inequality. Hence any experimental implementation requires a rigorous state preparation phase. For 3-dimensional systems there also exists a completely state independent test involving thirteen projectors developed by Oh and Yu \cite{Yu} and an almost state independent test involving nine projectors \cite{Kurzynski}. The latter is capable of revealing contextuality in every state, except the maximally mixed state, provided we choose the measurement basis appropriately. We describe this inequality in more detail in the following section. It has been shown that thirteen rays is the minimum number of distinct projector directions necessary for a state independent test of non-contextual realism \cite{Cabello13}.

The purpose of this paper is to propose an experimental setup for testing the almost state independent inequality from Ref. \cite{Kurzynski}. The experimental set up will be along the lines of Lapkiewicz et al.'s KCBS inequality experiment \cite{Zeilinger}. This provides an ideal platform for examining the recent debate over whether Lapkiewicz et al.'s setup is a faithful experimental implementation of contextuality tests based on n-cycle computability relation graphs, see \cite{Zeilinger, Ahrens}. We present an adaptation of the circuit which makes this issue transparent and could even be used to check the level of faithfulness empirically. We will also discuss the minimum detector efficiency required to ensure the results are independent of any auxiliary assumptions about photons in undetected events.

\section{Results}

Consider a theory with $n$ observables, $A_1,\dots A_n$, where each observable, $A_i$, returns outcome $a_i$ with probability $p(A_i = a_i)$. This theory exhibits classical realism whenever there exists a joint probability distribution for the combined measurement outcomes $p(A_1 = a_1,\dots, A_n = a_n)$ such that the measurement statistics, $p(A_i = a_i)$, of each observable, $A_i$, can be obtained as a marginal of this distribution \cite{Fine}. In this scenario there exists a non-contextual hidden variable model, conditioned on a set of parameters $\Lambda$. The joint probability distribution is reproduced through

\begin{equation}\label{eq:hiddenvariabledistribution}
p(A_1 = a_1,\dots, A_n = a_n) = \int ~ d\lambda ~ p(\lambda)p(A_1 = a_1| \lambda)\dots p(A_n = a_n| \lambda),
\end{equation}
 where each strategy for assigning measurement outcomes, $\lambda$, has weighted probability $p(\lambda)$ on the set $\Lambda$, and $p(A_i = a_i| \lambda)$ is the conditional probability of obtaining an outcome $a_i$ when measuring observable $A_i$ (for the given non-contextual hidden variable $\lambda$).

 Quantum theories do not necessarily exhibit non-contextual realism. When we measure an observable $A_1$ we simultaneously collapse the state we are observing onto an eigenstate of $A_1$.  If $A_1$ is independently simultaneously co-measurable with either member of a non-commuting pair ($A_2$ and $A_3$), then choosing to measure $A_1$ simultaneously with $A_2$ (respectively $A_3$) can erase information about the correlations between $A_1$ and $A_3$ (respectively $A_2$). Because a joint measurement of $A_1$ and $A_2$ can erase different information from a joint measurement of $A_1$ and $A_3$, the two joint measurements do not need to be unitarily equivalent. The choice to simultaneously co-measure $A_1$ with either  $A_2$ or $A_3$ can influence the outcome of $A_1$. We say that these two non-commuting observables each furnish a context for, and is compatible with, $A_1$ and that quantum mechanics is a contextual theory.

\begin{figure}[Htpb]
\vspace{-0.6cm}
\begin{center}
\includegraphics[width=0.4\textwidth]{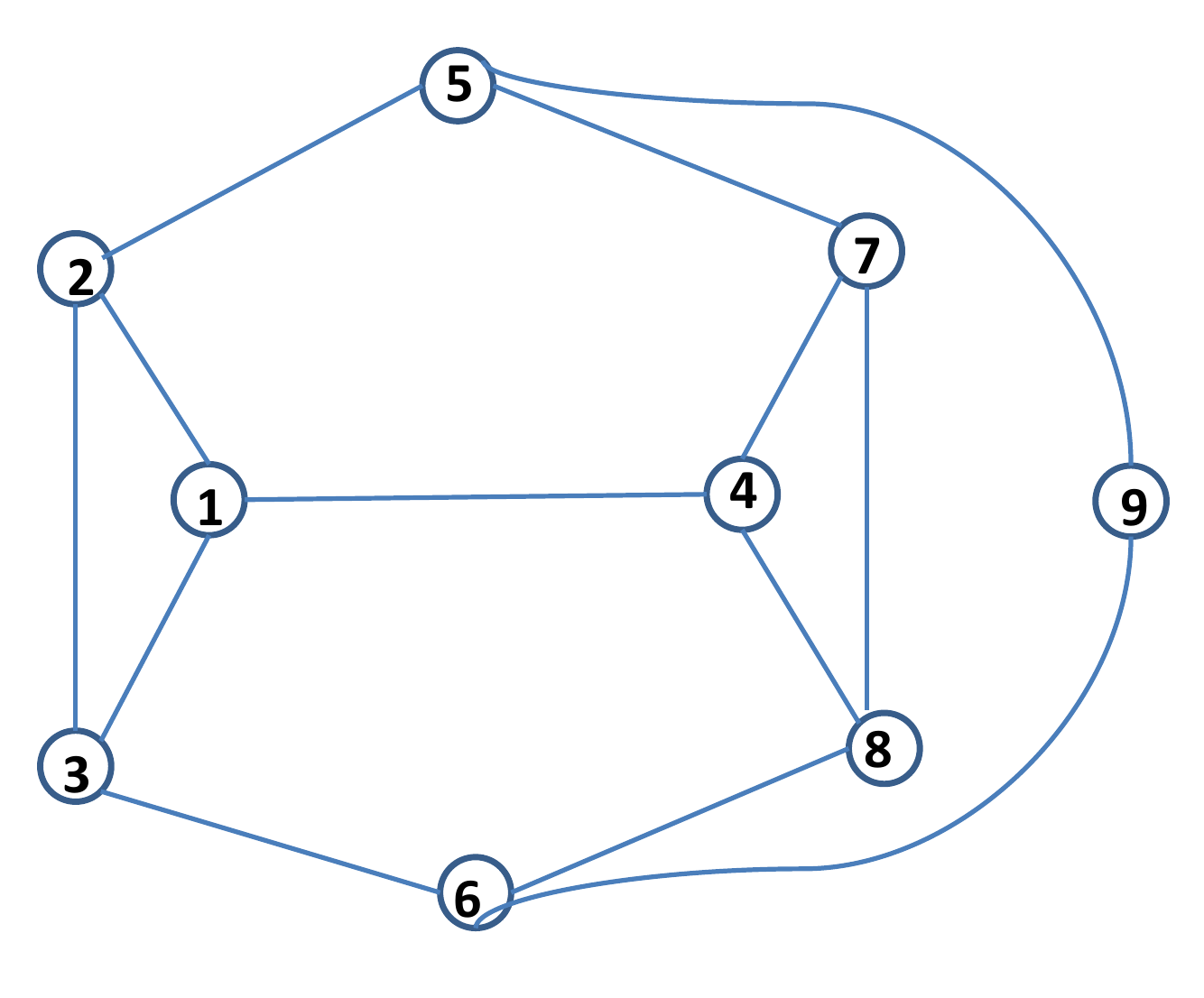}
\end{center}
\caption{The graph, $G$, of the computability relations for the nine observables, $A_1,\dots, A_9$, in the inequality \ref{dagandpawelcorrelationineqaulity} which is originally from Ref. \cite{Kurzynski}.}
 \label{fig:dagandpawelsgraph}
\end{figure}

Given a set of observables we can succinctly represent which subsets are compatible in a single graph. This graph has one vertex, $v_i$, for each observable $A_i$. The edge set of the graph is generated by connecting any two vertices $v_i$ and $v_j$ associated with co-measurable observables, respectively $A_i$ and $A_j$, by an undirected edge $(v_i, v_j)$.

As a concrete example consider the graph in Figure \ref{fig:dagandpawelsgraph} which outlines the compatability relations for a collection of nine observables in a three level system. These nine dichotomic $\{+1, -1\}$-observables, $A_1, \dots, A_9$, can be associated with projective measurements along nine rays in a 3-dimensional Hilbert space according to the relation:
\begin{equation}\label{eq:dichotomus}
A_i = 1 - 2 |i\rangle \langle i|.
\end{equation}
By construction these observables obey an exclusivity relation; for any set of mutually compatible observables $A_i, A_j, A_k$  the outcomes $A_i = -1$, $A_j = -1$ and $A_k = -1$ are exclusive. In other words when a measurement is made on a mutually commuting set of observables at most one observable will have outcome $-1$. If we try to preassign noncontextual outcomes to the nine observables in Figure \ref{fig:dagandpawelsgraph} which obey these exclusivity relations then we find the preassigned values always satisfy:

\begin{equation}\label{dagandpawelcorrelationineqaulity}
\sum_{(i,j) \in E(G)} \langle A_i A_j \rangle +\langle A_9\rangle  \ge -4.
\end{equation}

  However it was demonstrated in Ref. \cite{Kurzynski} that given a quantum state $\rho$, which is not the maximally mixed state, it is possible to choose a basis for the 3-dimensional Hilbert space so that Inequality (\ref{dagandpawelcorrelationineqaulity}) will be violated. More explicitly, the state $\rho$ violates Inequality (\ref{dagandpawelcorrelationineqaulity}) when rays used to define the observables $A_1,\dots, A_9$  are expressed in this basis \cite{Kurzynski}. Optimal violation will be obtained when we choose to measure with respect to the eigenbasis of $\rho$. To complement this analysis we have run a numerical simulation, using QI Mathematica package \cite{Miszczak}, which effectively tested Inequality (\ref{dagandpawelcorrelationineqaulity}) for a prefixed measurement basis and a randomly generated sample of ten million density matrices weighted by the Hilbert-Schmidt measure. We found $ 49.98\% $ of quantum states $\rho$ violated (\ref{dagandpawelcorrelationineqaulity}). A more detailed derivation of Inequality (\ref{dagandpawelcorrelationineqaulity}) is given in the Methods. In the following section we outline a proposal for experimentally testing Inequality (\ref{dagandpawelcorrelationineqaulity}).

\subsection{Experimental design}

We propose an experimental test of Inequality (\ref{dagandpawelcorrelationineqaulity}) for an indivisible 3-level quantum system furnished by a single photon. A photonic qutrit uses two distinct spatial paths for the photon and within one of the spatial paths two optical modes propagate as distinct polarization modes.
The core component is a scheme for simultaneous measurements of pairs of compatible observables $A_i$ and $A_j$ when the corresponding vertices $v_i$ and $v_j$ are connected by an edge in Figure \ref{fig:dagandpawelsgraph} which uses ideas from Lapkiewicz et al.'s recent experiment \cite{Zeilinger}. An implicit caveat on any experimental implementation of this proposal is that all experimental runs must be independent and the results after a statistically significant number of runs must represent an unbiased sampling from the joint probability distribution (\ref{eq:hiddenvariabledistribution}); in the case where the latter is tailored to the nine projective measurement constructed from (\ref{eq:nineprojectors}). Later we will discuss how losses and detector inefficiencies may allow a joint probability distribution conditioned on a hidden variable to violate Inequality (\ref{dagandpawelcorrelationineqaulity}). In both scenarios we use a single photon heralded source which will alow us to monitor photon losses. Previous experiments, \cite{Zeilinger, Zu}, produce pairs of polarization entangled photons in the singlet state via parametric down conversion and subsequently rerouted one photon to an auxiliary detector `detector $0$'. Post selecting on whether this auxiliary detector clicks creates a heralded single photon, see the Supplementary material for a detailed schematic. Any desired initial state can be synthesized using polarizing beam splitters to combine the two spatial modes and half wave plates to transform the polarization components inside a single spatial mode.

We introduce a single detector for each mode (we will use the label: detector $i$, for $i = 1,2 \textrm{ or } 3$). Two scenarios are considered: in the first scenario there are no losses and detector $i$ will click whenever a photon is in the corresponding mode $|i\rangle$. In the second scenario we will allow for inefficient detectors. We revisit this scenario later. It is convenient to identify the three orthogonal modes with rays $|1\rangle, |2\rangle $ and $|3\rangle $ from Equation (\ref{eq:nineprojectors}).  If detector $i$ clicks then we record $\Pi_i  = 1$ and using Equation (\ref{eq:dichotomus}) we assign a value $ A_i  = -1$ to the corresponding dichotomous observable.

The correlations of two dichotomous observables are given by:
\begin{eqnarray}\label{eq:correlations}
\langle A_i A_j\rangle &=& P(A_i = A_j =1)+ P(A_i = A_j = -1)  - P(A_i = 1 , A_j = -1) - P(A_i = -1, A_j = 1) \notag\\
&=& 1- 2P(A_i = 1 , A_j = -1) - 2P(A_i = -1, A_j = 1). \end{eqnarray}
The second line follows because (with unit probability) either both $A_i$ and $A_j$ are equal to $1$ or alternatively precisely one of them is equal to $-1$. We only count results from experimental runs where a single detector clicks coincidentally with detector $0$. This implies that in all events contributing to Equation (\ref{eq:correlations}) the sum of the measurement outcomes for the dichotomous observables (during a single run) is equal to $-1$ (in the first instance $ A_1 +  A_2 +  A_3 = -1$). Furthermore when $A_i$ and $A_j$ correspond to orthogonal  modes we discount experimental runs where $A_i=A_j = -1$ causing $P(A_i = A_j = -1) = 0$. In these circumstances it is pertinent that the projectors $|1\rangle$, $ | 2 \rangle$ and $|3 \rangle $ in Equation (\ref{eq:nineprojectors}) form a complete basis for the 3-dimensional Hilbert space. When sampling only from events where a single detector clicks, the measured correlations:
\begin{equation}\label{eq:cycles}
\langle A_1 A_2\rangle +\langle A_2 A_3\rangle + \langle A_3A_1\rangle = -1,
\end{equation}
obey the completeness relation. Under these experimental circumstance it is optional whether the above correlations need to be measured. A similar relation holds for the complete basis $|7\rangle $, $|4 \rangle$ and $|8\rangle$. We clarify that this completeness relation is not a fundamental assumption for deriving Equation (\ref{dagandpawelcorrelationineqaulity}), rather it is an optional simplification of the experiment which is available because these types of experiments post select on events where a single detector clicks coincidentally with detector $0$. This is a natural assumption in most experiments.

Repeated runs of this experimental configuration can be used to measure the correlations $\langle A_1 A_2\rangle$, $\langle A_1 A_3\rangle $ and $\langle A_2 A_3\rangle$. A full experiment capable of measuring all the correlation in Inequality (\ref{dagandpawelcorrelationineqaulity}) can be obtained by adding standard optical elements; specifically half wave plates and polarizing beam splitters. If we pass the input modes $|1\rangle$, $|2\rangle$ and $|3\rangle$ through a series of half wave plates and polarizing beam splitters then we can output the orthogonal modes for any three rays $|i\rangle, |j\rangle $ and $|k\rangle$ which are a linear combination of $|1\rangle, |2\rangle $ and $|3\rangle $, and form a basis for the 3-dimensional Hilbert space. At every stage of the experiment each detector is aligned to detect photons in a single mode. It is assumed that this detector will click if the photon is in the corresponding mode.

\begin{figure}[hbt!]
  \begin{center}
 \includegraphics[width=0.8\columnwidth]{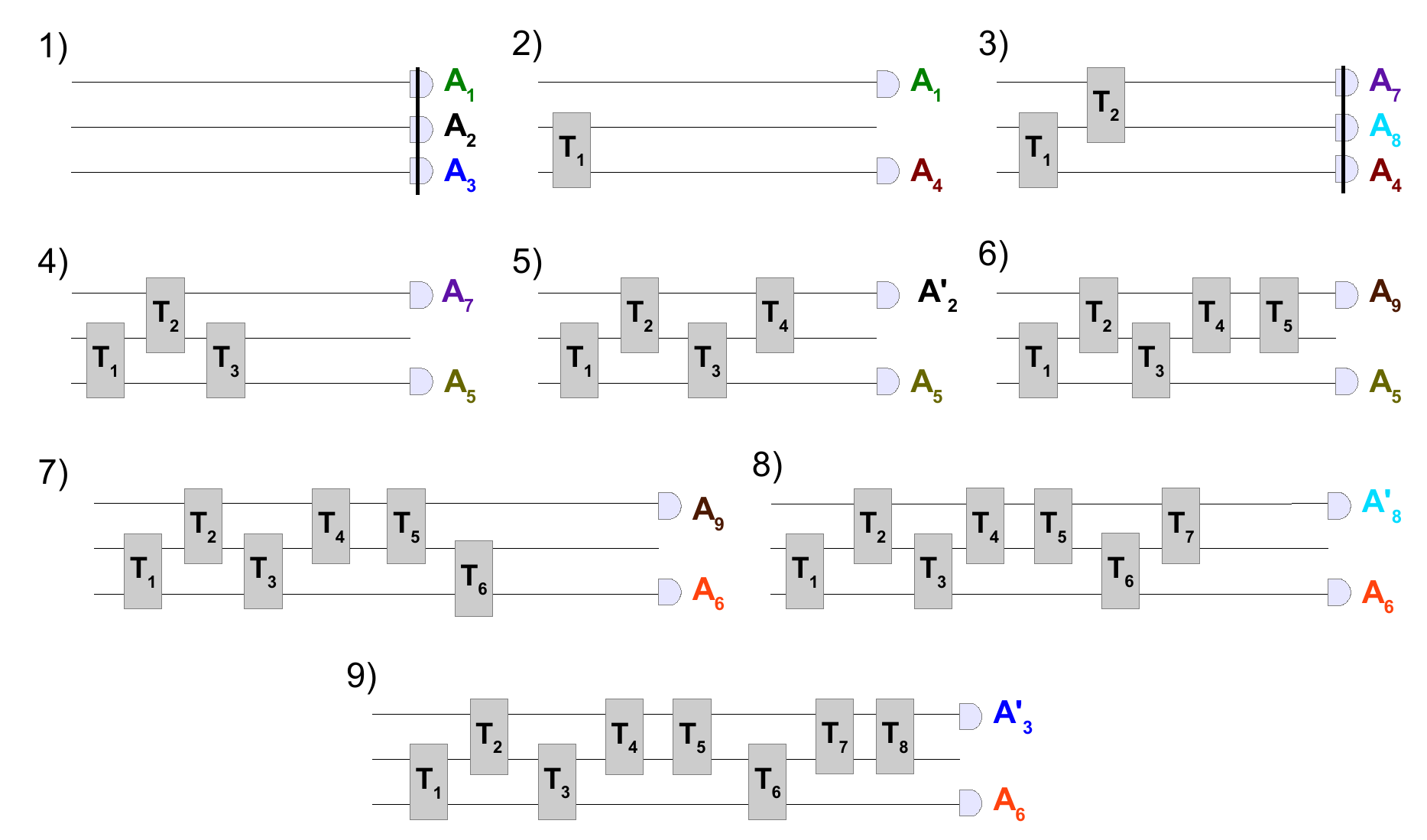}
  \end{center}
\caption{Each subfigure represents an experimental configuration which measures $\langle A_i A_j\rangle$ for the specific $A_i$ and $A_j$ indicated on the righthand side of the subfigure. Orthogonal modes are represented by horizontal lines. Sequences of half wave plates and polarizing beam splitters are labeled $T_1, \dots, T_8$ and detector positions are indicated by the corresponding dichotomous observable. The expectation value $\langle A_9\rangle $ can be obtained from the data collected during the experimental runs depicted in subfigures 6) or 7). The vertical line through the detectors in subfigures 1) and 3) indicate it is not necessary to record data during these stages. This follows from Equation (\ref{eq:cycles}). The precise sequence of optical elements $T_1, \dots, T_8$ are recorded in the Supplementary material. In this schematic when two consecutive stages of the experiment involve measuring the same mode, $|i\rangle$, and the beam line of mode $|i\rangle$ is unobstructed between the points at which these two measurements are made, we assume the physical implementation will measure the (same) mode $|i\rangle$ in two different contexts.}
  \label{fig:schematic}
\end{figure}

A schematic of the sequence of optical elements and measurements needed to obtain all the correlations in Inequality (\ref{dagandpawelcorrelationineqaulity})  is depicted in Figure \ref{fig:schematic}. The physical implementation of this experiment will require exact details regarding the sequence of half wave plates and  polarizing beam splitters needed to appropriately mix the two different polarization modes within a single spatial mode and subsequently mix the spatial modes. We leave the nuances of the setup to the discretion of an experimental group. However in the supplementary material we give a more detailed account of an explicit realization as a proof of principle that the schematic given in Figure \ref{fig:schematic} is physically realizable using an adaptation of Lapkiewicz et al.'s design \cite{Zeilinger}.

It is implicitly assumed that at any stage of the experiment; if a given mode's beam line is not interrupted by an optical element then there is perfect transmission of any photon  (amplitude component) in this mode. This occurs regardless of whether the other two modes pass through a half wave plate or are obstructed.

Notice that according to Figure \ref{fig:schematic}, we first measure the mutual correlations of the dichotomous observables, $A_1, A_2, A_3,$ corresponding to orthogonal modes $|1\rangle$, $|2\rangle$ and $|3\rangle$. Subsequently in subfigure 5) we need to measure  $\langle A_2 A_5\rangle $, however during the intermediary stages the beam line of mode $|2\rangle$ has been interrupted. This introduces a source of error. We must recreate the mode $|2\rangle$ before measuring $\langle A_2 A_5\rangle$. We can not be certain that in both instances we are measuring the precisely the same orthogonal mode $|2\rangle$. Correspondingly we introduce a new label $A_2'$ to distinguish between the observable $A_2$ in subfigure 1) and the observable $A_2'$ measured in subfigure 5) of Figure \ref{fig:schematic}.

In order to find the correlation term $\langle A_2 A_5\rangle$ we set up the experiment with the intention of measuring $A_2'$ however we create a time delay, $\Delta t$, for all photons in mode $|2\rangle$ by increasing the optical path length at the point where mode $|2\rangle$ is first created. This will ensure that photons originally in mode $|2\rangle$ will reach a detector $\Delta t$ seconds later than the click of the heralding detector, $D_0$. Whereas photons originally in modes $|1\rangle$ and $|3\rangle$ will result in detection events where $D_0$ clicks coincidently with one of the other three detectors. We now can calculate the correlation term $\langle A_2 A_5\rangle$ using Equation (\ref{eq:correlations}) and assigning outcome $A_2 = 1$ and $A_5 = -1$ to all runs where detector $5$ clicks coincidently with the auxiliary detector $D_0$, and analogously letting outcome $A_2 = -1$ and $A_5 = 1 $ correspond to all runs where one of the other two detectors clicks $\Delta t$ seconds later than the auxiliary detector $D_0$. This procedure generalizes to measuring the correlations $\langle A_3 A_6\rangle$ and $\langle A_8 A_6\rangle$, see the schematic \ref{fig:schematic}. We highlight that the physical implementation of this scheme will involve changing the (path length of the) beam line depicted in subfigure 5) of Figure  \ref{fig:schematic}. In theory any operation preformed on a photon originally created in mode $|2\rangle $ will have no effect on the outcome of the compatible observable $A_5$; in practice imperfect reconstruction of mode $|2'\rangle$ (and the orthogonal mode $|5\rangle$) may lead to violation of the no-disturbance principle. As a result there should be a careful analysis of the number of events where detector $5$ clicks $\Delta t$ seconds later than the heralding detector. These events where the photon is recoded to have been in both modes $|2\rangle$ and $|5\rangle$, equivalently $A_2 = A_5  = -1$, indicate a violating of the exclusivity condition.

Finally we highlight that Equation (\ref{eq:cycles}) implies we do not need to collect data during the experimental stages depicted in subfigures 1) and 3) of Figure \ref{fig:schematic}. We have drawn a vertical line through the detectors whenever it is not necessary to collect data.  We include these stages in the schematic because it is important to comprehensively set up all the measurement configurations. We need to ensure the orthogonal modes have the same relative orientations (and orthogonality relations) as the projectors in Equation (\ref{eq:nineprojectors}). This makes it essential to setup all stages of the experiment.

Errors will arise from imperfect alignment of half wave plates. This will cause some of the modes to be be misaligned with the rays in Equation (\ref{eq:nineprojectors}). In this case we will be testing Inequality (\ref{dagandpawelcorrelationineqaulity}) using projective measurement along directions which do not perfectly match the list in Equation (\ref{eq:nineprojectors}). The nine direction in Equation (\ref{eq:nineprojectors}) corresponded to a theoretically optimal set of projective measurements for testing Inequality (\ref{dagandpawelcorrelationineqaulity}). The errors introduced by half wave plate misalignment will make it more difficult to experimentally violate this inequality. However they will not invalidate any experimental results which directly demonstrate violation.

\section{Discussion}

We highlight that instead of measuring the correlation term $\langle A_2A_5\rangle$ using the time delay of the photon to register the event $A_2 = -1$, we could have modified Inequality (\ref{dagandpawelcorrelationineqaulity}),
\begin{equation}\label{eq:errortermineq}
\sum_{(i,j) \in E(G)} \left[\langle A_i A_j \rangle (1 - \delta_{i5}\delta_{j2} - \delta_{i,6} \delta_{j8} - \delta_{i6}\delta_{j3})\right] + \langle  A_5 A_2'\rangle + \langle  A_6 A_8'\rangle + \langle  A_6 A_3'\rangle  +  \langle A_9\rangle  \ge -1 - \epsilon_a - \epsilon_b - \epsilon_c,
\end{equation}
to include the error terms $\epsilon_a  = 1- \langle A_2 A_2'\rangle, ~ \epsilon_b = 1- \langle A_8 A_8'\rangle, ~ \epsilon_c = 1- \langle A_3 A_3'\rangle$. This would allow us to treat the correlation term $\langle A_2' A_5\rangle$ identically to all other previous measurements. However the error term $\epsilon_a$ would require us to redo the above analysis of $\langle A_2A_5\rangle$ in the almost identical case of $\langle A_2 A_2'\rangle$; again using the time delay of the photon to register an event where $A_2  = -1$ and the clicks of detector $2'$ to record events where $A_2' = - 1$. After collecting the data we can choose to use it to evaluate Inequality (\ref{eq:errortermineq}) and/or Inequality (\ref{dagandpawelcorrelationineqaulity}); in practice these two experiments are equivalent up to postprocessing of the data. This approach of measuring the modified Inequality (\ref{eq:errortermineq}) closely mirrors ideas from Lapkiewicz et al.'s experiment \cite{Zeilinger}. In this context it has been argued that experimentally testing Inequality (\ref{eq:errortermineq}) does not constitute a proper test of a noncontextual inequality \cite{Ahrens}. The controversy arises because  mode $|2\rangle$ which is originally measured in the context of modes $|1\rangle$ and $|3\rangle$ must be reconstructed before we measure the correlation term $\langle A_2' A_5\rangle$; this has instigated a debate on whether experimental implementations of Inequality (\ref{eq:errortermineq})  do measure precisely the same observable $A_2$ in two different contexts. In the spirit of Lapkiewicz et al.'s experiment \cite{Zeilinger} the condition for claiming observables $A_2$ and $A_2'$ are the same measurement in two different contexts is that they return identical empirical results; if they do not the error term $\epsilon_a$ will compensate. We note that using this approach the data collected (and therefore correlations measured) is identical no matter how it is post processed; and that if there is still ambiguity then we can evaluate both Inequalities to determine whether they are empirically equivalent.

 Any implementation of this proposal faces more critical problems from the key assumption that results collected during this experiment represent an unbiased sampling from the joint probability distribution (\ref{eq:hiddenvariabledistribution}). This requires us to make subsidiary assumptions about the states of undetected photons unless the detection efficiency is about a minimum thresh hold value $82 \% $. A detailed calculation is given in the next section.

Under these conditions, the above procedure could be used to experimentally test of the inequality from Ref. \cite{Kurzynski}. This implementation would probe one of the foundational principles distinguishing quantum mechanics from classical realistic descriptions of the world.  Furthermore if this test reveals contextual behavior in an indivisible spin one system then the behavior can not stem from entanglement. This result complements work done on the KCBS inequality \cite{Klyachko, Zeilinger} as well as the thirteen projector state independent inequality from Ref. \cite{Zu, Yu}. We present numerical studies which indicate Inequality (\ref{dagandpawelcorrelationineqaulity}) is violated by $49.98\% $ of states in a 3-level system. This makes an experimental test less sensitive to state preparation.

\section{Methods}

In this section we revisit the mathematical details required to derive Inequality (\ref{dagandpawelcorrelationineqaulity}) and the detector efficiency threshold needed to beat the no-detection loophole in this experiment.

Inequality (\ref{dagandpawelcorrelationineqaulity}) is based on nine projective measurements whose compatibility relations are provided by orthogonality and depicted in Figure \ref{fig:dagandpawelsgraph}. The nine vertices in Figure  \ref{fig:dagandpawelsgraph} can be identified with projective measurements $\Pi_i = |i\rangle\langle i|$, along the following nine rays in a 3-dimensional Hilbert space:
 \begin{eqnarray}\label{eq:nineprojectors} |1\rangle&=&(1,0,0), \nonumber \\
|2\rangle&=&(0,1,0), \nonumber \\ |3\rangle&=&(0,0,1), \nonumber \\
|4\rangle&=&1/\sqrt{2}(0,1,-1), \nonumber \\
|5\rangle&=&1/\sqrt{3}(1,0,-\sqrt{2}), \nonumber \\
|6\rangle&=&1/\sqrt{3}(1,\sqrt{2},0), \nonumber \\
|7\rangle&=&1/2(\sqrt{2},1,1), \nonumber \\
|8\rangle&=&1/2(\sqrt{2},-1,-1), \nonumber \\
|9\rangle&=&1/2(\sqrt{2},-1,1).  \end{eqnarray}

In a non-contextual theory, for any set of orthogonal rays $\{|i\rangle, |j\rangle, \dots \}$ it is possible to assign values of $1$ or $0$ to the corresponding projectors $\{\Pi_{i}, \Pi_j, \dots \}$, in such a way that the outcomes $\Pi_i  =1$ and $\Pi_j = 1$ are exclusive and if a projector, $\Pi_{i}$, appears in two different orthogonal sets then it is assigned the same consistent value in each set \cite{Peres}.

It follows that for any non-contextual realistic theory, where measurement statistics can be encapsulated in a hidden variable strategy $p(\lambda)$ as in (\ref{eq:hiddenvariabledistribution}), these nine-projectors satisfy:
\begin{equation}\label{eq:dagandpawelprojectorineq}
\sum_{i=1}^9 \langle \Pi_i\rangle  \, \le \,  3.
\end{equation}

The exclusivity relations of the dichotomic observables defined in Equation (\ref{eq:dichotomus}) are inherited directly from the exclusivity relations of the corresponding projectors, $\Pi_i = |i\rangle \langle i|$. Hence the principles and principles and assumptions behind Inequalities (\ref{dagandpawelcorrelationineqaulity}) and ({\ref{eq:dagandpawelprojectorineq}}) are equivalent. We can derive Inequality (\ref{dagandpawelcorrelationineqaulity}) directly from this result, by using Equation (\ref{eq:dichotomus}) to replace each of the projectors in Inequality (\ref{eq:dagandpawelprojectorineq}) by the corresponding dichotomous observable.

We note that any violation of the exclusivity conditions may lead to higher violations of Inequality (\ref{dagandpawelcorrelationineqaulity}) \cite{Guhne, Markiewicz, Budroni}.

These inequalities are violated by all quantum state $\rho$ except the maximally mixed state, provided we choose the basis for the 3-dimensional Hilbert space appropriately.

\subsection{Detector inefficiency}

A key assumptions in our proposed experimental test of Inequality (\ref{dagandpawelcorrelationineqaulity}) was that the results collected during a statistically significant number of runs represented an unbiased sampling from the joint probability distribution (\ref{eq:hiddenvariabledistribution}). In the presence of no-detection events however this assumption needs to be re-examined.

In any realistic experiment there will be runs where the photon is lost and none of the detectors click. These losses are caused by reflections off half wave plates and detector inefficiencies. In the proposed experimental scheme we only collect data from experimental runs where a single detector clicks; this automatically discounts any non-detection events. To claim the results of the proposed experiment represent an unbiased sampling from the joint probability distribution (\ref{eq:hiddenvariabledistribution}) we must assume the probability of loosing a photon is not correlated with the measurement outcome. In other words we must assume something about the state of the photon we never measured. Here we describe how efficient the experiment must be in order to eliminate this assumption.

 Without loss of generality we can assume all photons are lost at the detector. We attribute an efficiency $\eta$ to each of the three detectors.  We want to make our results independent of the distribution of photon modes in undetected runs. To do this we include the runs where none of the detectors click. The combined total number of runs is constituted by runs where either a single detector clicks or none of the detectors click.

 In the ideal scenario where there are no losses, $p(A_i = a_i)$ is the probability of getting outcome $a_i$ when you measure observable $A_i$. When we include losses we use $p'(A_i = a_i) = \eta p(A_i = a_i)$ for the proportion of the combined total number of runs where we measure $A_i$ and get outcome $a_i$.

 To uniquely specify the results of a run we must specify the outcomes $a_i, a_j $ and $a_k$ of all three dichotomous observables (respectively $A_i$, $A_j$ and $A_k$).

When estimating the correlations in Equation (\ref{eq:correlations}) we now use the renormalized probability:
\begin{equation}
\widetilde{p}(A_i = a_i, A_j = a_j, A_k = a_k) = \frac{p'(A_i = a_i, A_j = a_j, A_k = a_k)}{1- p'(A_i=A_j=A_k  =1)},
\end{equation}
which is the ratio of the number of runs with the given outcome ($ A_i = a_i, A_j = a_j, A_k = a_k$)  to the number of runs where a single detector clicks. This eliminates the dependence on no detection events (runs with outcome $A_i= A_j =A_k = 1$).

In hidden variable theories we can use (\ref{eq:hiddenvariabledistribution}) to express $\widetilde{p}(A_i = a_i, A_j = a_j, A_k = a_k)$ as a function of the detector efficiency $\eta$ and probabilities in the ideal scenario where there are no losses:
\begin{equation}
\widetilde{p}(A_i = a_i, A_j = a_j, A_k= a_k) =\frac{1}{1- (1-\eta)} \int d\lambda ~ p(\lambda)~\eta  p(A_i = a_i |\lambda) ~ \eta  p(A_j = a_j| \lambda) ~ \eta p(A_k = a_k|\lambda).
\end{equation}
This allows us to evaluate the correlations in Equation (\ref{eq:correlations}) for a hidden variable strategy $p(\lambda)$ conditioned on $\lambda$. We find Inequality (\ref{dagandpawelcorrelationineqaulity})  is modified in the presence of detector inefficiency to
\begin{equation}\label{eq:modifiedSKIinequality}
\sum_{(i,j) \in E(G)} \langle A_i A_j \rangle +\langle A_9\rangle  \ge -4/\eta^2.
\end{equation}
We need $\eta > 0.82$ for the right hand side of Equation (\ref{eq:modifiedSKIinequality}) to be below the maximum level of quantum violation for this inequality \cite{cirelson}. Hence when the detector efficiency is below $82 \%$ it is impossible to experimentally detect violation of Inequality (\ref{dagandpawelcorrelationineqaulity}) unless subsidiary assumptions are made about the  distribution of putative outcomes (which detector failed to click) for undetected events.

\section{Acknowledgements}
\begin{acknowledgements}
This work was supported by the National Research
Foundation and Ministry of Education in Singapore. PK is also supported by the Foundation for Polish Science. We thank Valerio Scarani and Rafael Rabelo for useful discussions on the experimental set up. JT thanks both Su-Yong Lee and Mile Gu for discussion about the practicality of physically implementing the above scheme. RP thanks Dagomir Kaszlikowski and Centre for Quantum Technologies
for hosting his visit to Singapore and nurturing his budding film career.
\end{acknowledgements}

\section{Supplementary material}

\begin{figure}[hbt!]
  \begin{center}
 \includegraphics[width=0.8\columnwidth]{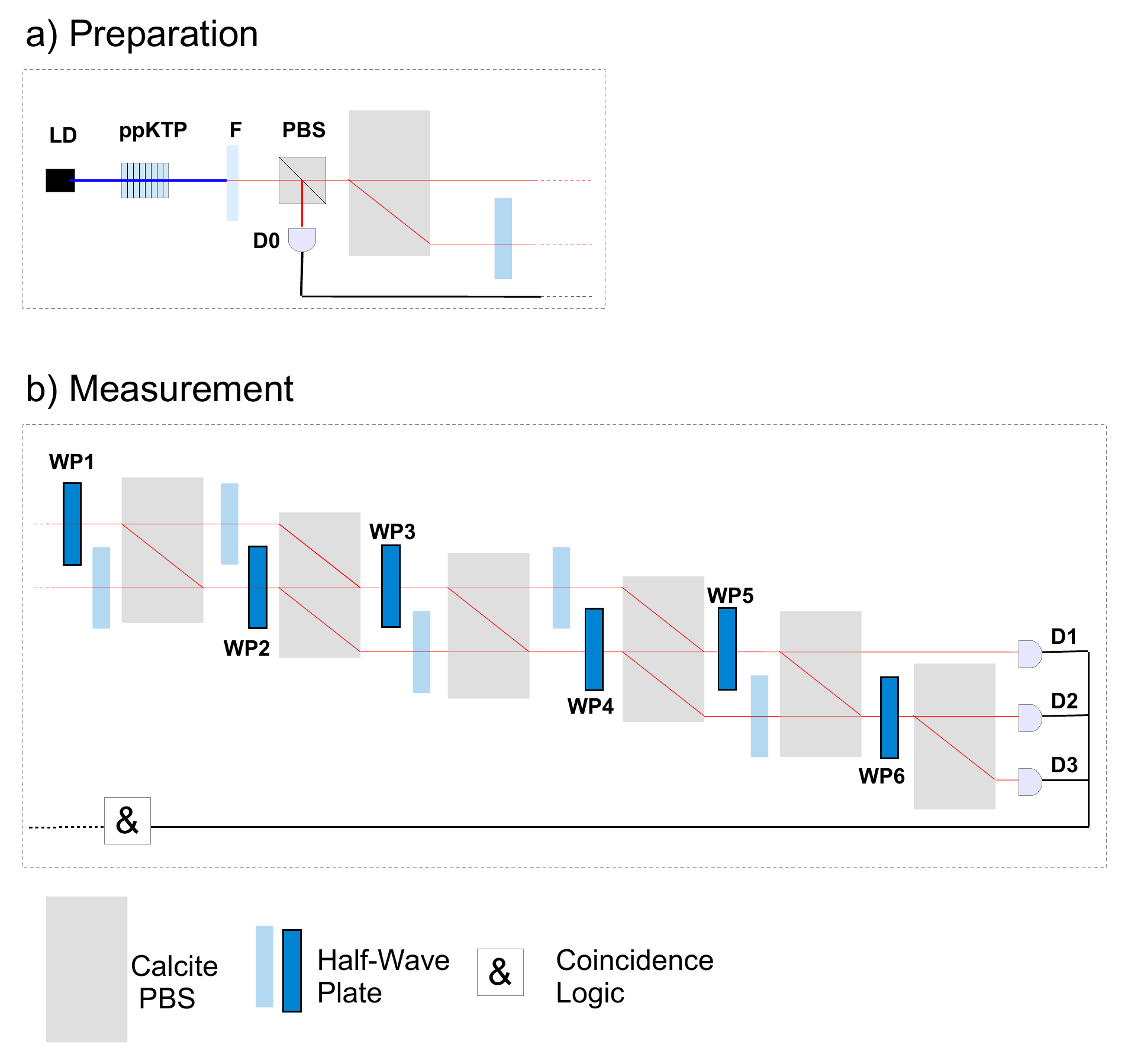}
  \end{center}
\caption{Here we give a possible experimental implementation of Schematic 2 in the spirit of Lapkiewicz et. al. [5]. {\bf Preparation}: a single photon heralded source is created through parametric down conversion; a process which outputs pairs of polarization entangled photons in the singlet state. Subsequently a polarizing beam splitter is used to redirect one member of the pair to a heralding detector: `detector 0' which is denoted by `{\bf 0}' in the figure. Post selecting on events where detector 0 clicks clearly identifies runs where the photon is lost (detector 0 is the only detector to click) as distinct from runs where the photon is not lost (one of detectors 1-3 click coincidentally with detector 0).  We take inspiration from Ref. [5] which used a 405 nm Laser diode (LD) in combination with a nonlinear-periodically-poled-Potassium-Titanyl-Phosphate-crystal (ppKTP) and suitable filters (F) to produce pairs of polarization entangled photons. For a comprehensive explanation of this process see Ref. [5]. Any desired initial state can now be synthesized using additional half wave plates and polarizing beam splitters. {\bf Measurement}: there are always two concurrent modes with the same spatial index and opposite (orthogonal) polarization indices. Half wave plates $WP1-WP6$ manipulate these two modes.  Each wave plate implements a $2 \times 2$ rotation on the two dimensional subspace spanned by the pair of polarization modes within a single spatial path. The angles characterizing these rotations are given in Table \ref{tab:halfwaveplates}. Modes with different spatial indices can be recombined using polarizing beam splitters which are represented by grey rectangles. We use $0$ radians to denote the configuration of an optical element (alignment of the optical axis) which implements the identity transformation. The remaining light-blue wave plates have been inserted to balance the path lengths of the two spatial modes; these are all set to $\pi/4$ (this alignment balances a $\pi$ phase shift between the two spatial modes). When the run requires a subset of the optical elements, the unnecessary elements can be aligned to identity transformations. Note that in Figure 2 transformations $T_4$ and $T_5$ act consecutively on the same two-mode subspace. In the physical implementation $T_4$ and $T_5$ correspond to wave plate $WP4$ set to two different orientations, see Table \ref{tab:halfwaveplates}. Analogously $T_7$ and $T_8$ use two different settings of $WP6$.}
  \label{fig:opticaltable}
\end{figure}

As a proof of principle that the ideas present in this paper are practically realizable we give a more comprehensive account of how to implement Schematic 2. More detailed information about the sequence of optical elements needed to implement the experiment is given in the following figure. This figure should be read in combination with the accompanying table which lists of transformations that half wave plates $WP1-WP6$ enact on the two polarization modes. Each angle corresponds to the angle of rotation in a $2 \times 2$ rotation matrix acting on a pair of polarization modes within a single spatial path. We have chosen to give the angle characterizing the transformation rather than the precise configuration of the wave plate's optical axis because the latter will be contingent on the exact details of the experimental implementation and may be obsolete if even a small change is made. The `$-$' indicates
that the corresponding wave plate is unnecessary for measuring the corresponding correlation.
\newline

\begin{tabularx}{\textwidth}{ |X|X|X|X|X|X|X| }
\hline
\label{tab:halfwaveplates}
{\bf Correlations measured} & $WP1$ & $WP2$ &$WP3$ & $WP4$  & $WP5$ & $WP6$  \tabularnewline \hline
$\langle A_1 A_2\rangle $  $\langle A_1 A_3\rangle$  $\langle A_2 A_3\rangle$ & - & - & - &- &- &- \tabularnewline \hline
$\langle A_1 A_4\rangle$ & $\frac{3 \pi}{4}$ & - & - &- &- &- \tabularnewline \hline
$\langle A_7 A_8\rangle $  $\langle A_8 A_4\rangle$  $\langle A_7 A_4\rangle$ &  $\frac{3 \pi}{4}$ &  $\frac{ \pi}{4}$  & - &- &- &-\tabularnewline \hline
$\langle A_5 A_7\rangle$ & $\frac{3 \pi}{4}$ & $\frac{ \pi}{4}$ & $\textrm{cos}^{-1}(\frac{1}{\sqrt{3}})$&- &- &- \tabularnewline \hline
$\langle A_2' A_5\rangle$ & $\frac{3 \pi}{4}$ & $\frac{ \pi}{4}$ & $\textrm{cos}^{-1}(\frac{1}{\sqrt{3}})$ & $\frac{\pi}{3}$ &- &- \tabularnewline \hline
$\langle A_9 A_5\rangle$ & $\frac{3 \pi}{4}$ & $\frac{ \pi}{4}$ & $\textrm{cos}^{-1}(\frac{1}{\sqrt{3}})$ & $\frac{13\pi}{12}$ &- &- \tabularnewline \hline
$\langle A_9 A_6\rangle$ & $\frac{3 \pi}{4}$ & $\frac{ \pi}{4}$ & $\textrm{cos}^{-1}(\frac{1}{\sqrt{3}})$ &$\frac{13\pi}{12}$ &$\textrm{cos}^{-1}(\frac{1}{3})$ &- \tabularnewline \hline
$\langle A_8' A_6\rangle$ & $\frac{3 \pi}{4}$ & $\frac{ \pi}{4}$ & $\textrm{cos}^{-1}(\frac{1}{\sqrt{3}})$ & $\frac{13\pi}{12}$  &$\textrm{cos}^{-1}(\frac{1}{3})$ & $-\frac{\pi}{3}$ \tabularnewline \hline
$\langle A_3' A_6\rangle$ & $\frac{3 \pi}{4}$ & $\frac{ \pi}{4}$ & $\textrm{cos}^{-1}(\frac{1}{\sqrt{3}})$ & $\frac{13\pi}{12}$ &$\textrm{cos}^{-1}(\frac{1}{3})$ & $\frac{\pi}{3}$\tabularnewline \hline
\end{tabularx}
\newline

\section{Author contributions}

J.T., R.P., P.K. and D.K. designed the proposed experiment and provided theoretical analysis. R.P. ran the mathematica simulation and designed figures. J.T. and P.K. wrote the manuscript. D.K. supervised the project, edited the manuscript and provided original ideas and direction.

\section{Competing financial interests}
The authors declare no competing financial interests.


\begin{thebibliography}{99}
\bibitem{Footnote1}
These preassigned outcomes (for all observables $A_1,\dots, A_5$) must be independent of which consecutive pair of observables $A_i$ and $A_{(i + 1  \,{\rm mod } \, 5)}$ you intend to measure.
\bibitem{Footnote2}
Quantum nonlocality can be reinterpreted as specialized case of quantum contextuality, making tests like the CHSH inequality \cite{CHSH} the minimal test of quantum contextuality.
\end{thebibliography}
\end{document}